\documentclass[floatfix,aps,showpacs,pra,twocolumn]{revtex4}
\usepackage{graphicx}

\begin{document}

\newcommand{\x}{{\bf x}}
\renewcommand{\k}{{\bf k}}
\newcommand{\E}{{\bf E}}
\newcommand{\e}{{\bf e}}
\def\mn#1{\marginpar{*\footnotesize \footnotesize #1}{*}}
\def\isum{{\sum\!\!\!\!\!\!\!\!\int}}
\def\bfzero{{\bf 0}}
\newcommand{\ket}[1]{\, | #1 \rangle}
\newcommand{\be}{\begin{equation}}
\newcommand{\ee}{\end{equation}}
\newcommand{\bea}{\begin{eqnarray}}
\newcommand{\eea}{\end{eqnarray}}
\newcommand{\bdm}{\begin{displaymath}}
\newcommand{\edm}{\end{displaymath}}
\newcommand{\bfd}{{\bf d}}
\newcommand{\bfk}{{\bf k}}
\def\hF{{\bar F}}
\newcommand{\bfp}{{\bf p}}
\newcommand{\bfq}{{\bf q}}
\newcommand{\bfE}{{\bf E}}
\newcommand{\bfP}{{\bf P}}
\newcommand{\ha}{{\hat a}}
\newcommand{\bfr}{{\bf x}}
\newcommand{\bfz}{{\bf 0}}
\newcommand{\cE}{{\cal E}}
\newcommand{\tG}{{\widetilde G}}
\newcommand{\veps}{\varepsilon}
\newcommand{\non}{\nonumber \\}
\def\pder#1#2{\frac{\partial #1}{\partial #2}}
\def\lsim{\raise0.3ex\hbox{$<$\kern-0.75em\raise-1.1ex\hbox{$\sim$}}}

\title{Dynamics of matter-wave and optical fields in superradiant scattering 
from Bose-Einstein condensates}
\author{O.~Zobay and Georgios~M.\ Nikolopoulos}
\affiliation{Institut f\"ur Angewandte Physik, Technische Universit\"at
Darmstadt, 64289 Darmstadt, Germany}
\date{\today}
\begin{abstract}
We study superradiant scattering off Bose-Einstein condensates by solving 
the semiclassical Maxwell-Schr\"odinger equations describing the coupled dynamics 
of matter-wave and optical fields. Taking the spatial dependence of these 
fields along the condensate axis into account, we are able to reproduce 
and explain many of the characteristic features observed in the experiments of Inouye {\it et al.} [Science {\bf 285}, 571 
(1999)] and Schneble {\it et al.} [Science {\bf 300}, 475 
(2003)], such as the shape of the atomic side-mode distributions for forward 
and backward scattering, the spatial asymmetry between forward and 
backward side modes, and the depletion of the condensate center observed for forward scattering.
\end{abstract}
\maketitle

\flushbottom
 
The recent observation of superradiant scattering from Bose-Einstein condensates (BECs) \cite{InoChiSta99,SchTorBoy03} has significantly extended our knowledge about collective emission processes.
BEC superradiance differs in several key aspects from the 
well-known ``conventional" superradiance involving electronically excited atoms \cite{Dic54,GroHar82}. In particular, the role of the excited state is now played by the combination of a BEC and an impinging laser pulse, spontaneous emission is replaced by spontaneous Rayleigh scattering, and the ground state corresponds to atoms in well-defined momentum side modes \cite{InoChiSta99,SchTorBoy03}.

The experimental observation of BEC superradiance
was initially interpreted in the framework of matter-wave stimulation \cite{InoChiSta99}. In this 
picture, interference between the recoiling atoms and the condensate at 
rest leads to a matter-wave grating, from which laser 
photons are scattered. This causes the matter-wave grating to grow rapidly in a collective, self-amplifying process.
However, later experiments with shorter and stronger 
laser pulses \cite{SchTorBoy03} led to a more profound understanding of BEC superradiance and showed that the above picture was 
incomplete. 
In these latter experiments, backward-scattered atoms were observed in 
addition to the forward peaks. As suggested by the MIT group, the new
experimental results should not be explained in terms of optical diffraction 
from a matter-wave grating, but as atomic diffraction from the optical 
grating formed by the superposition of the impinging laser beam and the 
Rayleigh-scattered photons. An essential cornerstone of their argumentation 
was the fact that forward- and backward-scattered atoms show a spatial 
asymmetry. It was conjectured that this asymmetry arises because atomic diffraction is strongest at the edges of the condensate, 
where the intensity of the optical grating is largest. 
On the other hand, for pure forward scattering the picture of matter-wave 
stimulation is equally valid, which is evidenced by condensate depletion 
occurring mainly at the center where the matter-wave grating is supposed 
to be most pronounced.

Subsequently, however, this interpretation of asymmetric 
scattering was questioned by Meystre and co-workers. 
According to \cite{PuZhaMey03}, the observed asymmetry may also be due to the 
fact that larger angles are favored in backward scattering, because of a 
reduced energy mismatch. 
Nevertheless, their theoretical analysis was not able to 
distinguish between the two alternatives. It is one of the main purposes of 
the present paper to resolve this controversy. 
As our results show, the explanation put forward by the MIT group is indeed 
the correct one, thus confirming the picture of optical stimulation. 

Our work is based on the semiclassical solution of the spatially dependent 
Maxwell-Schr\"odinger equations for the coupled optical and matter-wave 
fields. In this way, we are able to overcome two main limitations of the theory of Ref.\ \cite{PuZhaMey03}, i.e., the restriction to short times and first-order scattering (undepleted-pump approximation) as well as the neglect of propagation effects. 
So far, the latter have also been disregarded in most other theoretical treatments of BEC superradiance, e.g., \cite{VasEfiTri04,RobPioBon04}. References \cite{AveTri04,BonPioRob05} discuss spatial effects in the BEC-light interaction, but do not provide a detailed comparison to the results of Refs.\ \cite{InoChiSta99,SchTorBoy03}.
The present results show that the 
inclusion of propagation effects is essential for a comprehensive theoretical 
understanding of BEC superradiance.
Indeed, 
our model enables us to also reproduce and explain many other characteristic features, such as the typical momentum distribution patterns in forward 
and backward scattering and the depletion of the condensate center in 
forward scattering. All these results clearly show that our model indeed 
captures the essential aspects of the physics of BEC superradiance in both the weak- and strong-pulse regimes. 
Thus, whereas the MIT interpretation stresses the difference of 
the physical pictures used to explain the observations in the two regimes, our approach emphasizes the existence of a unifying 
theoretical framework.

In our theoretical treatment, we consider a cigar-shaped condensate oriented along the $z$ axis. 
The BEC is exposed to a linearly polarized laser pulse 
$\cE_0(t){\bf e}_y (e^{i(k_lx-\omega_l t)}+{\rm c.c.})/2$, $\omega_l = ck_l$, 
travelling in the $x$ direction. The laser is far off-resonant from the 
atomic transition to the excited electronic state $\ket e$. After 
adiabatically eliminating the state $\ket e$, the coupled 
Maxwell-Schr\"odinger equations of motion for the mean-field macroscopic 
wave function $\psi(\x,t)$ and the positive- and negative-frequency 
components $\bfE^{(\pm)}(\x,t)$ of the classical electric field read
\cite{GroHar82,ZhaWal94}
\bea\label{psi_ad}
i\hbar\pder{}t\psi &=& -\frac{\hbar^2}{2M}\Delta\psi+\frac{(\bfd\cdot\bfE^{(-)})(\bfd\cdot\bfE^{(+)})} {\hbar\delta}\psi,\\
\label{E_field}
\pder{^2 \bfE^{(\pm)}}{t^2} &=& c^2 \Delta \bfE^{(\pm)}-\frac 1{\veps_0} \pder{^2 \bfP^{(\pm)}}{t^2}
\eea
with $\delta$ the detuning of the electronic transition, $\bfd$ the atomic dipole moment and $M$ the atomic mass. The polarisation is given by
$
\bfP^{(+)}(\x,t)= -\bfd |\psi(\x,t)|^2 \bfd\cdot\bfE^{(+)}(\x,t) /\hbar\delta
$, $\bfP^{(-)}= \bfP^{(+)\,*}$.
Note that in Eq.\ (\ref{psi_ad}) we neglect the external trapping potential 
and the atomic interactions, since they do not play a significant role on the time scales of 
the process under consideration. However, it would be straightforward to 
include them in the model. 

We will solve Eqs.\ (\ref{psi_ad}) and (\ref{E_field}) under the slowly-varying-envelope approximation (SVEA). To this end, we decompose the fields as
\bea\label{svea_psi}
\psi(\x,t) &=& \sum_{(n,m)} \frac{\psi_{nm}(z,t)}{\sqrt A} e^{-i(\omega_{n,m}t-nk_lx - mkz)},\\ \label{svea_ep}
\E^{(+)}(\x,t) &=& \cE_0\e_y e^{-i(\omega_l t-k_lx)}/2 +\cE_+(z,t)\e_y e^{-i(\omega t-kz)}\non
&& + \cE_-(z,t)\e_y e^{-i(\omega t+kz)},
\eea
$\E^{(-)}(\x,t)=\E^{(+)\, *}(\x,t)$, with $\omega=kc$ and $A$ the average condensate cross section perpendicular to the $z$ axis. 
The electric field in Eq.\ (\ref{svea_ep}) contains the impinging laser pulse together with 
the two optical endfire modes, that are produced by collective Rayleigh 
scattering. The endfire modes, whose envelope functions are denoted $\cE_\pm(z,t)$, travel up and down the condensate axis, respectively.
For simplicity, we model the applied laser pulse as rectangular lasting 
from $t=0$ up to $t=t_f$. In Eq.\ (\ref{svea_psi}), the summation is over 
all momentum side modes $(n,m)$, with $m+n$ even. In the side mode $(n,m)$, 
atoms possess momentum $\hbar(nk_l\e_x + mk\e_z)$ and have a slowly varying 
spatial envelope $\psi_{nm}(z,t)$, while their kinetic energy is given by 
$\hbar\omega_{n,m} = \hbar^2(n^2k_l^2+m^2k^2)/2M$. In this notation, 
the ``side mode" $(0,0)$ describes the condensate at rest. 
The wave vector $k$ is fixed by energy conservation for 
the transitions between the side modes $(0,0)$ and $(1,\pm 1)$ which 
initiate the process, i.e., $\hbar ck_l = \hbar ck +\hbar \omega_{1,1}$. 
Since $k_l-k \ll k,k_l$, we can approximate $\omega_{n,m} \approx (n^2+m^2)\omega_r$ 
with the recoil frequency $\omega_r = \hbar k_l^2/2M$. We also introduce 
$\Delta \omega = \omega-\omega_l=-2\omega_r$ and 
$\omega_{n,m}^{\pm\pm}=\omega_{n,m}-\omega_{n\pm 1,m\pm 1}\approx -2(\pm n\pm m +1)\omega_r$.
Note that in the ansatz (\ref{svea_psi})-(\ref{svea_ep}) we disregard the dependence of the envelope functions $\psi_{nm}$ and 
$\cE_\pm$ on the transverse directions $x$ and $y$. For the matter waves, 
this is certainly a good approximation since the radial degrees of freedom 
are tightly confined by the trap. For the optical fields, we can use this 
approximation, because the Fresnel number of the system is close to 1 
\cite{InoChiSta99,GroHar82}.

Using the ansatz (\ref{svea_psi})-(\ref{svea_ep}) and introducing the 
rescaled fields $e_\pm = \cE_\pm/\sqrt{\hbar\omega/2\veps_0 A}$, 
Eq.\  (\ref{psi_ad}) in the SVEA reads
\bea\label{env_psi}
&&i\pder{\psi_{nm}} t =
-\frac\hbar{2M} \pder{^2 \psi_{nm}}{z^2} 
-i \frac {m\hbar k}{M} \pder{\psi_{nm}}z\non
&&+g\sqrt{L} 
\left[ e_+^*\psi_{n-1,m+1}e^{i(\Delta\omega+
\omega_{n,m}^{-+}) t} 
+e_-^*\psi_{n-1,m-1} e^{i(\Delta\omega+\omega_{n,m}^{--}) t}\right.\non 
&&\left .+e_+\psi_{n+1,m-1}e^{i(-\Delta\omega+
\omega_{n,m}^{+-}) t}+e_-\psi_{n+1,m+1} e^{i(-\Delta\omega+\omega_{n,m}^{++}) t} \right]
\eea
with the coupling constant \cite{RobPioBon04}
\bdm
g=\frac{|\bfd|^2\cE_0}{2\hbar^2\delta} \sqrt{\frac{\hbar\omega}{2\veps_0 AL}}
\edm
and the condensate length $L$.
The first term on the right-hand side of Eq.\ (\ref{env_psi}) describes the 
quantum-mechanical dispersion of the envelope function, while the second one 
leads to a spatial translation with velocity $v_m = m\hbar k/M$. 
The other terms 
describe the spatially dependent light-induced couplings of the momentum side mode $(n,m)$ to other modes. In particular, through stimulated scattering, an atom in a side mode $(n,m)$ can absorb a laser photon and deposit it into one of the endfire modes. The accompanying recoil transfers the atom into one of the side modes 
$(n+1,m\pm 1)$. Alternatively, the atom may absorb an endfire-mode photon and emit it into the laser beam, thereby ending up in the side mode 
$(n-1,m\pm 1)$. This latter process is responsible for atomic backward 
scattering \cite{SchTorBoy03}.
Finally, following \cite{SchTorBoy03}, we disregard absorption and emission 
between endfire modes. 

Neglecting retardation effects, the envelope functions $e_\pm$ are given by
\bea\label{e_p}
e_+(z,t) &=& -i\frac{g\sqrt L}c\int_{-\infty}^z dz' \sum_{(n,m)} e^{i(\Delta\omega -\omega_{n,m}^{+-}) t}\non
&& \times\psi_{nm}(z',t)\psi_{n+1,m-1}^*(z',t),\\
\label{e_m}
e_-(z,t) &=& -i\frac{g\sqrt L}c\int^{\infty}_z dz' \sum_{(n,m)} e^{i(\Delta\omega -\omega_{n,m}^{++}) t}\non
&& \times\psi_{nm}(z',t)\psi_{n+1,m+1}^*(z',t).
\eea 
From these equations one sees how the build-up of the endfire-mode fields 
is driven by the coherences $\psi_{nm}\psi_{n+1,m\pm 1}^*$.

The semiclassical Eqs.\ (\ref{env_psi})-(\ref{e_m}) describe the evolution of the system in the macroscopic regime of superradiance where the populations of atomic and optical modes are large compared to one. As discussed in Refs.\ \cite{GroHar82,HaaKinSch79}, such equations have to be solved with stochastic initial conditions (which we will call ``seeds") that model the random character of the initial quantum noise starting up the process. In our case, the noise is due to spontaneous Rayleigh scattering transferring condensate atoms into the side modes $(1,\pm 1)$ \cite{MooMey99}. For any given initial condition, the corresponding solution of the semiclassical equations describes one possible realization of the experiment. The variations between different simulations reflect the macroscopic effects of the initial quantum fluctuations. We have solved Eqs.\ (\ref{env_psi})-(\ref{e_m}) for a variety of seed functions. We find that for fixed external parameters, despite the differences in quantitative details, all solutions share characteristic features, such as the side-mode distribution patterns. This shows that these features originate from specific semiclassical dynamical mechanisms and not from quantum fluctuations, and it is on these semiclassical effects that we will focus in the following.

In the examples shown below, we use the seed function $\psi_{1,\pm 1}(z,0) = \psi_{0,0}(z,0)/\sqrt N$ for concreteness (with $N$ denoting the atom number). This corresponds to having one delocalized atom in each of the side modes. Furthermore, following the experimental data of Ref.\ \cite{SchTorBoy03}, we consider a $^{87}$Rb BEC with $N=2\times 10^6$ and length $L=200\,\mu$m. The condensate is in the Thomas-Fermi regime, so that we can model its wave function as $\psi_{0,0}(z,0) = \sqrt{n(z)}$ with $n(z) = C[(L/2)^2-z^2]\Theta(L/2-|z|)$, $C=3N/4L^3$. The pulse duration $t_f$ and the coupling strength $g$ are in the same regime as those used in the experiments (compare with Ref.\ \cite{RobPioBon04}).

\begin{figure}[t]
\centerline{\includegraphics[width=8.0cm]{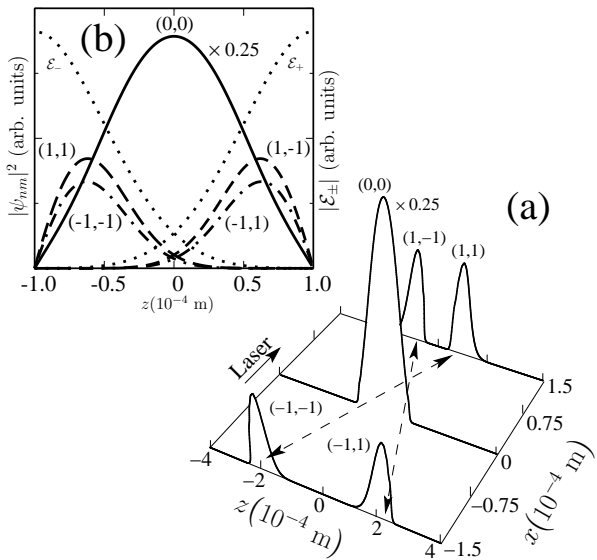}}
\caption{Strong-pulse regime. 
(a) Spatial distribution of the first-order forward $(1,\pm 1)$ and 
backward $(-1,\pm 1)$ 
atomic side modes, after applying a laser pulse of duration 
$t_f=14\,\mu$s and strength $g=2\times 10^6\,$s$^{-1}$ to the condensate 
followed by a free propagation for a time $t_p=25\,$ms.
(b) Spatial distributions of the atomic side modes and the optical 
endfire modes $({\cal E}_\pm)$, at time $t_f$. For the sake of 
illustration the BEC population $(0,0)$ has been divided by $4$.
\label{fig1}}
\end{figure}

Let us first consider the regime of strong laser pulses, which is characterized by the superradiant gain \cite{InoChiSta99} being much larger than the recoil frequency $\omega_r$.
In Fig.\ \ref{fig1}(a), we display a snapshot of the atomic spatial distribution after applying a strong laser pulse to the condensate followed by a subsequent 
free propagation for a time $t_p\gg t_f$. Since we work with a one-dimensional 
model, we calculate the displacement $\Delta x$ between the condensate and the first-order side modes in the $x$ direction as $\Delta x=v_r t_p$ with the recoil velocity $v_r=\hbar k/M=5.9\times 10^{-3}\,$m$/$s. Our result clearly reproduces the asymmetry observed in Fig.\ 2 of Ref.\ \cite{SchTorBoy03}. Moreover, we are able to verify the explanation for this phenomenon conjectured in \cite{SchTorBoy03}: As depicted in Fig.\ \ref{fig1}(b), at time 
$t_f$ the atomic side modes and the optical field modes are indeed localized near the condensate edges, as suggested by the MIT group. Related results were also obtained in a simpler model in Ref.\ \cite{AveTri04} disregarding backward scattering. The reason for the behavior shown in Fig.\ \ref{fig1} can be inferred from the linearized analysis of Eqs.\ (\ref{env_psi})-(\ref{e_m}) \cite{ZobNik05}. One finds that, during the build-up of the atomic side modes, the electric fields $\cE_\pm (z,t)$ grow steadily in the $z$ and $-z$ direction, respectively, and are strongest around the ends of the condensates. Equation (\ref{env_psi}) then implies that the side modes will predominantly grow in these areas of large electric fields where the atomic diffraction is largest [see Fig.\ \ref{fig1}(b)]. The observed asymmetry is now due to the fact that during the subsequent free time evolution, the backwards scattered atoms immediately travel further outwards, i.e., away from the condensate center. The forward scattered atoms, however, will initially move inwards  [dashed lines in Fig.\ \ref{fig1}(a)].
 
Inspecting Figs.\ 1(A) and 2 of \cite{SchTorBoy03}, we also find evidence against the interpretation of Pu, Zhang, and Meystre \cite{PuZhaMey03}. According to their Fig.\ 3(b), forward peaks should predominantly appear at an angle of 45$^\circ$ from the condensate center; the experiment, however, clearly shows smaller angles (towards the direction of the laser pulse). Their Fig.\ 3(b) also predicts backward scattering predominantly at 45$^\circ$, but with angles larger than 45$^\circ$ favored compared to those less than 45$^\circ$. In the experiments there is no backward scattering at 45$^\circ$, but only at larger angles. Furthermore, Ref.\ \cite{PuZhaMey03} predicts that the superradiant scattering always involves atoms with momenta ${\bf q}$ and $-{\bf q}$ simultaneously. It is thus not clear how atoms can appear at different angles in the forward and backward directions. The underlying reason for the failure of the model of Ref.\ \cite{PuZhaMey03} is the neglect of spatial propagation effects, since our theory shows that these are crucial to understand the experimental observations. 

\begin{figure}[t]
\centerline{\includegraphics[width=7.0cm]{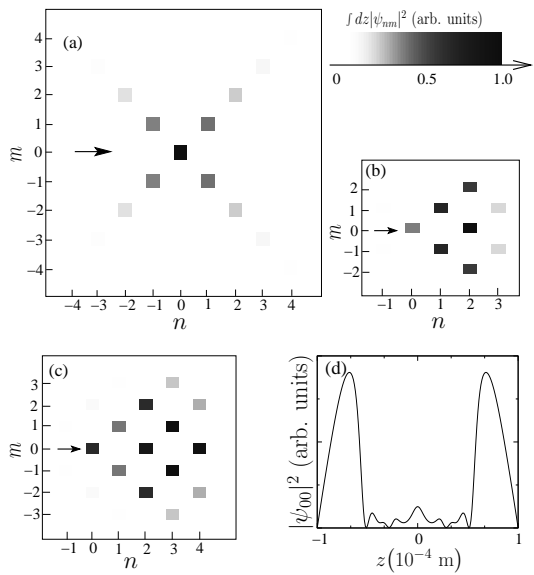}}
\caption{Atomic side-mode distributions.  Each square represents an 
  integrated probability $p_{nm}= \int dz |\psi_{nm}(z,t)|^2$. 
  (a) Strong-pulse regime: $t_f=10.6\,\mu$s and $g=2.6\times 10^6\,{\rm s}^{-1}$.
  (b) Weak-pulse regime:  $t_f=232\,\mu$s and $g=5.0\times 10^5\,{\rm s}^{-1}$.
  (c) Weak-pulse regime:  $t_f=291\,\mu$s and $g=6.5\times 10^5\,{\rm s}^{-1}$.
  (d) Spatial distribution of the condensate along the axis $z$ corresponding 
  to (c). 
\label{fig2}}
\end{figure}

In Fig.\ \ref{fig2}(a), we show the atomic side-mode distribution forming an 
X-shape pattern after applying a strong pulse. This figure should be compared to Fig.\ 1(A) of Ref.\ \cite{SchTorBoy03}, where the momentum distribution was obtained through time-of-flight imaging. In discussing this figure, we want to emphasize that the very appearance of an X-shape pattern, i.e., the suppression of off-diagonal side modes with $|n|\ne |m|$, is another independent confirmation of the fact that in the strong-pulse regime the atomic side modes and the optical endfire modes are located at the condensate edges. The off-diagonal side mode $(2,0)$, for example, is resonant with the modes $(1,\pm 1)$, but it can only be populated if the $(1,\pm 1)$ modes overlap with the endfire modes $\cE_\pm$, respectively 
[see Eq.\ (\ref{env_psi})]. Therefore, the absence of a peak for $(2,0)$ implies that there is no such overlap [see Fig.\ \ref{fig1}(b)]; i.e., the modes must be located near the edges. The side modes $(2,\pm 2)$, however, are readily populated since the side modes $(1,\pm 1)$ overlap with $\cE_\mp$. We want to stress that the X-shape pattern cannot be explained by models neglecting propagation effects, e.g., Refs.\ \cite{VasEfiTri04,RobPioBon04}.
In these models off-diagonal side modes such as $(\pm 2,0)$ and $(0,\pm 2)$ become strongly populated as well, so that an X pattern does not emerge. This again proves the significance of spatial effects in the strong-pulse regime.

We now turn to the weak-pulse regime in which only forward-scattered atoms 
with a characteristic fan-like side-mode pattern are observed 
(see Figs.\ 1(E)-(G) of \cite{InoChiSta99}, Fig.\ 1(B) of 
\cite{SchTorBoy03}, Fig.\ 2(a) of \cite{YoSuToKu04}). As evidenced in Figs.\ \ref{fig2}(b,c), our model is able to reproduce these patterns. We would like to emphasize that the 
appearance of such patterns is a typical, not a coincidental, outcome in our simulations.
We also find other types of distributions, such as the existence of population only on the forward diagonals $n=|m|$, $n>0$. Such patterns, which have not yet been reported experimentally, appear as 
an intermediate regime between the full X-shape and the fan pattern. 
As depicted in Fig.\ \ref{fig2}(d), our model also 
reproduces the depletion of the condensate center in the weak-pulse regime (see inset of Fig.\ 1(B) of \cite{SchTorBoy03}).
Our model predicts that the onset of the center depletion is correlated with the appearance of the $(2,0)$ side mode. This is due to the fact that the $(2,0)$ mode can only be populated efficiently if the $(1,\pm 1)$ modes overlap with the ${\cal E}_\pm$ endfire modes, respectively. This first happens around the center of the atomic sample, since both atomic and optical modes gradually expand from the edges towards to the center. At that point, the condensate, from which the $(1,\pm 1)$ modes grow, is depleted there, whereas it has been replenished at the edges due to Rabi flopping \cite{ZobNik05}.

The presented examples show that the atomic side-mode distributions form a sensitive probe providing detailed insight into the coupled dynamics of optical and matter-wave fields. A systematic investigation of these distributions, e.g., their dependence on laser pulse duration and strength, would constitute a form of time-resolved spectroscopy of BEC superradiance.

Besides studying characteristic semiclassical effects, our theory also allows us to examine macroscopic quantum fluctuations in BEC superradiance. Following Refs.\ \cite{GroHar82,HaaKinSch79}, we have investigated the delay fluctuations of the first maximum in the intensity of the emitted superradiant light. In both the weak- and strong-pulse regime the delay distributions are found to be well described by the Gumbel distribution familiar from ``conventional" superradiance \cite{GroHar82}.

In summary, we have presented an analysis of superradiant scattering from 
Bose-Einstein condensates in terms of the spatially dependent semiclassical Maxwell-Schr\"odinger equations. We have been able to reproduce and explain 
several characteristic features of the experiments, that have not been 
theoretically accounted for so far.
We finally note that our theory equally well applies to the recently demonstrated matter-wave amplification \cite{MWA}, and we expect it to be an important tool for a thorough understanding of this process as well.

\end{document}